\documentclass
[twocolumn,english,aps,prl,amssymb,floatfix,superscriptaddress,showpacs]{revtex4}%
\usepackage[T1]{fontenc}
\usepackage{amsmath}
\usepackage{graphicx}
\usepackage{amssymb}
\usepackage{graphicx}
\usepackage{graphicx}
\usepackage{amsfonts}
\usepackage{graphicx}
\usepackage{babel}%
\begin{document}
\title{Absence of conventional quantum phase transitions in itinerant systems with disorder}
\author{V. Dobrosavljevi\'{c}}
\affiliation{Department of Physics and National High Magnetic Field Laboratory, Florida
State University, Tallahassee, Florida 32306, USA.}
\author{E. Miranda}
\affiliation{Department of Physics and National High Magnetic Field Laboratory, Florida
State University, Tallahassee, Florida 32306, USA.}
\affiliation{Instituto de F\'{\i}sica Gleb Wataghin, Unicamp, Caixa Postal 6165, Campinas,
SP, CEP 13083-970, Brazil.}
\date{\today{} }

\begin{abstract}
Effects of disorder are examined in itinerant systems close to quantum
critical points. We argue that spin fluctuations associated with the
long-range part of the RKKY interactions generically induce non-Ohmic
dissipation due to rare disorder configurations. This dissipative mechanism is
found to destabilize quantum Griffiths phase behavior in itinerant systems
with arbitrary symmetry of the order parameter, leading to the formation of a
"cluster glass" phase preceding uniform ordering.

\end{abstract}

\pacs{75.40.-s, 71.10.Hf, 71.27.+a, 75.10.Nr}
\maketitle

The effects of quenched disorder seem to be especially important near quantum
phase transitions and lead to new classes of phenomena dominated by rare
disorder configurations. At present, this behavior is rather well understood
in insulating disordered magnets, leading to features such as the infinite
randomness fixed point and the associated quantum Griffiths phase anomalies
\cite{McCoy69,dsf9295,ThillHuse95,YoungRieger96,RiegerYoung96,gbh96,senthilsachdev,Pich98,Motrunich00}%
. Quantum critical regions, though, are much easier to access in itinerant
systems such as heavy fermion materials, where quantum fluctuations can be
tuned by adjusting the Fermi energy of the charge carriers.

In many of these systems, behavior reminiscent of quantum Griffiths phases
seems to arise, triggering much theoretical controversy and debate. One early
interpretation concentrated on the interplay of disorder and the Kondo effect
far from any magnetic ordering, leading to the {}\textquotedblleft electronic
Griffiths phase\textquotedblright\ \cite{miranda}. Another scenario assumed
that the phenomenon relates to rare regions close to magnetic ordering
\cite{CastroNetoJones}, similar to those in insulating magnets. More recent
work \cite{MillisMorrSchmalian}, however, shed some doubt on the validity of
the latter picture for itinerant systems. It emphasized the crucial role of
Landau damping, which was argued to suppress the low temperature Griffiths
phase behavior for magnets with Ising symmetry. In a remarkable paper
\cite{smearingprl}, Thomas Vojta discussed this mechanism on general symmetry
grounds, and demonstrated that in the Ising case dissipation leads to a
fundamental modification of the quantum critical point (QCP). Subsequent work
\cite{VojtaSchmalian} suggested that the same mechanism \textit{does not}
apply for other symmetry classes such as Heisenberg magnets, where the
standard QCP scenario is argued to hold, preceded by a conventional Griffiths phase.

\begin{figure}[ptb]
\begin{center}
\includegraphics[  width=3.4in,
keepaspectratio]{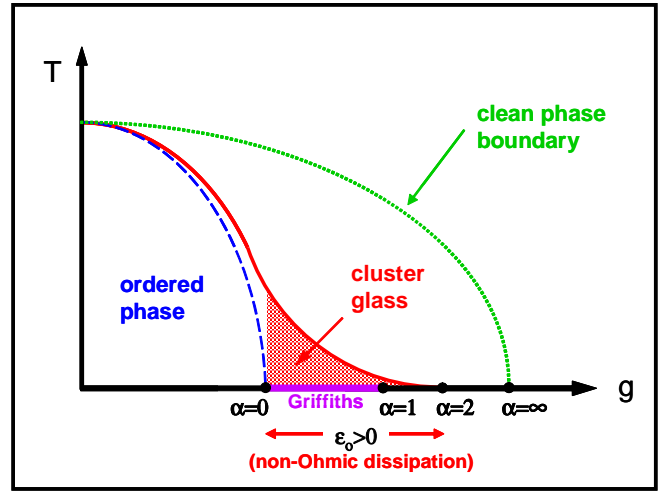}
\end{center}
\caption{Phase diagram of an itinerant system with disorder, close to the
clean quantum critical point (QCP). Shown are the clean critical line (dotted
green) and that in the presence of disorder if non-Ohmic dissipation is
ignored (dashed blue line). In this case, a Griffiths phase emerges close to
the magnetically ordered phase, for $\alpha<1$ (purple). When dissipation is
present, sufficiently large droplets {}\textquotedblleft
freeze\textquotedblright\ already for $\alpha<2$, leading to the formation of
the "cluster glass" phase (shaded region), preceding the uniform ordering. }%
\label{cap:fig1}%
\end{figure}

All these works concentrated on the dynamics of a \textit{single} droplet
(rare region) in an Ohmic dissipative environment. In this letter we argue
that this physical picture is incomplete. In itinerant systems, magnetic
moments - classical or quantum - are known \cite{MillisMorrSchmalian,narozhny}
to interact through \emph{long-ranged} RKKY interactions. As a result, even
the dynamics of spatially distant droplets cannot be considered as
independent. The dynamics of any droplet is therefore affected not only by
dissipation due to the conduction electrons in its vicinity, but also by spin
fluctuations due to RKKY interactions with many other droplets. This effect is
especially important in Griffiths phases where droplets are characterized by
power-law distributions of local energy scales, with a tunable dynamical
exponent $d/z^{\prime}$. In the rest of this letter we present a
self-consistent theory describing the dynamics of interacting droplets. Within
our theory: (1) dissipation produced by such spin fluctuations can acquire a
more singular non-Ohmic form which can adversely affect the dynamics for all
symmetry classes; (2) any Griffiths phase is destabilized by this mechanism,
leading to the generic formation of a ``cluster glass'' phase preceding any
uniform ordering, thus introducing a fundamental modification of the
conventional QCP scenario.

\textit{Model of interacting droplets.} To be specific, we concentrate on the
same regime as explored in Ref.~\cite{VojtaSchmalian}, and examine a weakly
disordered itinerant antiferromagnet near a QCP. In the absence of disorder
and within the standard Landau-Ginzburg-Wilson description, the distance to
the QCP is measured by the coupling constant $r.$ When disorder is present,
exponentially rare regions (droplets) can arise which are nearly magnetically
ordered. Within the $i$-th droplet of size $L_{i},$ the local value of the
(bare) coupling constant is assumed to be $r_{i}\ll0$, while outside the
droplets (the {}\textquotedblleft bulk\textquotedblright) we have a weakly
correlated metal ($r_{o}\gg0)$. As discussed in Ref.~\cite{VojtaSchmalian}, if
the interactions between the droplets are ignored, then the dynamics of each
droplet is determined only by its size. After appropriate coarse-graining,
such a droplet is described by an action of the form%
\begin{align}
S_{drop}[r_{i},\phi_{i}]  &  =\int_{o}^{\beta}d\tau d\tau^{\prime}\,\,\phi
_{i}(\tau)\Gamma_{i}(\tau-\tau^{\prime})\phi_{i}(\tau^{\prime})\nonumber\\
&  +\frac{u}{2N}\int_{o}^{\beta}d\tau\,\,\phi_{i}^{4}(\tau).
\label{eq:dropaction}%
\end{align}
Here, $\phi_{i}$ is an $N$-component ($N>1$) order parameter field describing
the $i$-th droplet, and $\Gamma_{i}(\tau)$ is the coarse-grained two-point
vertex whose Fourier transform is%
\begin{equation}
\Gamma_{i}(\omega_{n})=\left(  r_{i}+|\omega_{n}|\right)  .
\end{equation}
The local {}\textquotedblleft bare\textquotedblright\ coupling constant of
each droplet $r_{i}\approx r(L_{i})^{d}$ reflects its size (here $r<0$ is the
coupling constant of the clean magnet). Large droplets are exponentially rare
(probability $P(L_{i})\sim\exp\{-\rho(L_{i})^{d}\}$), correspondingly
$P(r_{i})\sim\exp\{-\rho r_{i}/r\}$, where $\rho$ is the volume fraction of
magnetic droplets . Now, each droplet maps onto a one-dimensional classical
Heisenberg chain with $1/\tau^{2}$ interactions and the corresponding energy
gap $\varepsilon_{i}\sim\exp\{\pi r_{i}/u\}$ \cite{smearingprl,VojtaSchmalian}%
. The resulting distribution function $P(\varepsilon_{i})\sim\varepsilon
_{i}^{\alpha-1}$, giving rise to standard Griffiths phase behavior. The
{}\textquotedblleft Griffiths exponent\textquotedblright\ $\alpha=d/z^{\prime
}$ is a non-universal function of parameters, which is expected
\cite{VojtaSchmalian} to decrease as the magnetic ordering is approached; for
$\alpha<1$, the average susceptibility $\chi\sim T^{1-\alpha}$ diverges as
$T\rightarrow0$.

In contrast to the situation of Ref.~\cite{VojtaSchmalian}, when a finite
concentration of droplets is present, they will interact. The action
describing the droplet-droplet interaction can be obtained by formally
integrating over the order parameter fluctuations of the {}\textquotedblleft
bulk\textquotedblright\ separating the droplets. Using the fact that in this
region the spin fluctuations are only weakly correlated ($r_{o}\gg0$), we can
formally integrate out all the degrees of freedom in the bulk, to obtain an
effective interaction coupling pairs \cite{threebody} of droplets%
\begin{equation}
S_{ij}=\int_{o}^{\beta}d\tau d\tau^{\prime}\,\,\phi_{i}(\tau)\chi_{o}%
(R_{ij},\tau-\tau^{\prime})\phi_{j}(\tau^{\prime}), \label{eq:interdrop}%
\end{equation}
where $R_{ij}\gg L_{i},\,L_{j\text{ }}$ is the distance between droplets $i$
and $j$. The interaction kernel $\chi_{o}(R_{ij},\tau-\tau^{\prime})$ is, in
fact, nothing but the non-local susceptibility of the weakly correlated
{}\textquotedblleft bulk\textquotedblright\ and as such can be accurately
obtained from standard Fermi liquid considerations (see, e.g., \cite{narozhny}%
). Its detailed form is complicated and model-dependent, but for our problem
only its long-distance and low frequency components are of importance, since
the droplets are spatially distant and display slow dynamics. We can thus
safely ignore its frequency dependence and consider only its asymptotic form,
which is essentially that of the RKKY interaction%
\begin{equation}
\chi_{o}(R_{ij},\tau-\tau^{\prime})\approx\frac{J_{ij}}{(R_{ij})^{d}}%
\delta(\tau-\tau^{\prime}).
\end{equation}

In a disordered metal, impurity scattering introduces random phase
fluctuations in the usual periodic oscillations of the RKKY interaction,
which, however, retains its power law form (although its \textit{average}
value decays exponentially \cite{jagannathan,narozhny}). Hence, such an
interaction acquires a random amplitude $J_{ij}$ of zero mean and variance
$<J_{ij}^{2}>=J^{2}$ \cite{jagannathan,narozhny}.

\textit{Field theory of interacting droplets.} Because of the long-ranged
interactions, one needs to describe the collective dynamics of the entire set
of such droplets. To do this, we formally average over the random interactions
$J_{ij}$ using standard replica methods, and the interaction term takes the
form
\begin{equation}
S_{ij}\approx-\frac{1}{2}\frac{J^{2}}{(R_{ij})^{2d}}\sum_{a,b=1}^{n}\int
_{o}^{\beta}d\tau d\tau^{\prime}\,\,\phi_{i}^{a}(\tau)\phi_{i}^{b}%
(\tau^{\prime})\phi_{j}^{a}(\tau)\phi_{j}^{b}(\tau^{\prime}),
\end{equation}
where $a,b=1,...,n\;(n\rightarrow0)$ are the replica indices. We then
introduce collective $Q$-fields by decoupling $S_{ij}$ through a
Hubbard-Stratonovich transformation to write the replicated partition function
as%
\begin{equation}
Z^{n}=\int D\phi DQ\exp\{-S_{eff}[\phi,Q]\},
\end{equation}
where the effective action%
\begin{align}
S_{eff}[\phi,Q]  &  =\frac{1}{2}\sum_{a,b=1}^{n}{\displaystyle \sum
\limits_{ij,\omega_{n}}}Q_{i}^{ab}(\omega_{n})\frac{J^{2}}{(R_{ij})^{2d}}%
Q_{j}^{ab}(\omega_{n})\nonumber\\
&  +{\displaystyle \sum\limits_{i}}\left\{  S_{drop}[r_{i},\phi_{i}%
]+S_{MC}[\phi_{i},Q]\right\}  ,
\end{align}
and the {}``mode-coupling'' term%
\begin{equation}
S_{MC}[\phi_{i},Q]=\sum_{a,b=1}^{n}{\displaystyle \sum\limits_{j,\omega_{n}}%
}\phi_{i}^{a}(\omega_{n})\phi_{i}^{b}(\omega_{n})\frac{J^{2}}{(R_{ij})^{2d}%
}Q_{j}^{ab}(\omega_{n}).
\end{equation}

\textit{Saddle point theory.} Since the droplets are exponentially rare, their
interactions are also very small. To describe the leading order corrections it
is sufficient to evaluate the $Q$-fields at the saddle point level and the
mode-coupling term assumes the form%
\begin{equation}
S_{MC}[\phi_{i},\Delta]=-\sum_{a,b=1}^{n}{\displaystyle\sum\limits_{\omega
_{n}}}\phi_{i}^{a}(\omega_{n})\Delta^{ab}(\omega_{n})\phi_{i}^{b}(\omega_{n}),
\end{equation}
with the {}\textquotedblleft cavity field\textquotedblright\ $\Delta
^{ab}(\omega_{n})$ satisfying the following self-consistency condition%
\begin{equation}
\Delta^{ab}(\omega_{n})={\displaystyle\sum\limits_{i}}\frac{J^{2}}%
{(R_{ij})^{2d}}\left\langle \phi_{i}^{a}(\omega_{n})\phi_{i}^{b}(\omega
_{n})\right\rangle _{i},
\end{equation}
where the expectation value is evaluated with respect to the local effective
action $S_{loc}[r_{i},\phi_{i}]=S_{drop}[r_{i},\phi_{i}]+S_{MC}[\phi
_{i},\Delta]$. If all the droplets were the same, this set of equations would
be \textit{identical} to those describing a metallic quantum spin glass
\cite{sachdev}. For interactions $J$ sufficiently large, such a model would
feature a conventional quantum critical point, where spin-glass ordering would
set in at $T=0$. A crucial new feature in our model is a broad power-law
distribution $P(\varepsilon)\sim\varepsilon^{\alpha-1}$ of local energy scales
characterizing the droplets in a Griffiths phase, which produces non-Fermi
liquid behavior for $\alpha<1$. We will see, however, that already for
$\alpha<2$, qualitatively new dynamics emerges, completely changing the
critical behavior of the model.

\textit{Instability of the Griffiths phase.} Assuming that all droplets are
well separated, the average interaction strength is very small. If we could
ignore the fluctuations in the droplet sizes, the system would be far from any
ordering. We thus concentrate on the paramagnetic phase of our model, where $
\left\langle \phi_{i}^{a}(\omega_{n})\phi_{i}^{b}(\omega_{n})\right\rangle
_{i}=\delta_{ab}\chi_{loc}(\varepsilon_{i},\omega_{n}).
$ Since the droplet sizes (i.e. their local energy scales $\varepsilon_{i}$)
are independent of their position, we find%
\begin{equation}
\Delta^{ab}(\omega_{n})=\widetilde{g}\delta_{ab\,}\overline{\chi_{loc}%
(\omega_{n})}=\widetilde{g}\delta_{\alpha b}\,\int d\varepsilon_{i}%
P(\varepsilon_{i})\chi_{loc}(\varepsilon_{i},\omega_{n}),
\end{equation}
where the RKKY coupling constant $\widetilde{g}=J^{2}\sum_{i}(R_{ij}%
)^{-2d}\sim nJ^{2}$, where $n$ is the density of droplets per unit volume.

The {}\textquotedblleft memory kernel\textquotedblright\ $\overline{\chi
_{loc}(\omega_{n})}$ describes additional dissipation caused by spin
fluctuations resulting from long-ranged RKKY interactions between the
droplets. Its quantitative form is obtained by self-consistently solving the
dynamics of each droplet in such a dissipative bath and then carrying out the
appropriate averaging procedure. In this letter, however, we limit our
attention to demonstrating the instability of the Griffiths phase with respect
to adding the RKKY interactions. To do this, begin with $\widetilde{g}=0$ and
imagine turning on infinitesimal RKKY interactions. To leading order, we
replace
\begin{align}
\overline{\chi_{loc}(\widetilde{g},\omega_{n})}  &  \approx\overline
{\chi_{loc}(0,\omega_{n})}\sim\int d\varepsilon_{i}\varepsilon_{i}^{\alpha
-1}[\varepsilon_{i}+|\omega_{n}|]^{-1}\nonumber\\
&  =\overline{\chi_{loc}(0)}-\tilde{\gamma}|\omega_{n}|^{1-\varepsilon_{o}%
}+O(|\omega_{n}|),
\end{align}
where the $\varepsilon_{o}=2-\alpha$, and $\tilde{\gamma}$ is a constant. Note
that the dissipation acquires a \textit{non-Ohmic} form for $\alpha<2$. As
anticipated in Ref.~\cite{smearingprl}, such non-Ohmic dissipation immediately
changes the dynamics of the droplets. For $\varepsilon_{o}>0$, the Heisenberg
chain \cite{kosterlitz76} that represents the droplet dynamics (whose
interactions decay with a power slower than $r^{-2}$) finds itself
\emph{above} its lower critical dimension! Accordingly, sufficiently large
droplets ($L>L_{c}(\varepsilon_{o})$) {}\textquotedblleft
freeze\textquotedblright, i.e. behave as classical (superparamagnetic) objects.

\textit{{}Cluster glass phase.\ }As soon as some of the droplets freeze, they
immediately order at a finite (albeit exponentially low) temperature, forming
a ``cluster glass'' phase. To estimate the ordering temperature, we first need
to determine the critical droplet size, given the anomalous dissipation
exponent $\varepsilon_{o}$. For small values of $\varepsilon_{o}$, the
corresponding critical exponents have been calculated \cite{kosterlitz76}
using an $\varepsilon$-expansion approach. For our purposes, it suffices to
determine the {}\textquotedblleft transition temperature\textquotedblright\ of
the equivalent Heisenberg chain. This task is adequately accomplished using
the large-$N$ approach. The corresponding $T=0$ self-consistency condition for
the critical coupling constant reads ($\Lambda$ is an ultraviolet cutoff)%
\begin{equation}
r_{i,c}=-u\int_{0}^{\Lambda}\frac{d\omega}{\pi}\frac{1}{\omega+\gamma
\omega^{1-\varepsilon_{o}}},
\end{equation}
where $\gamma=\tilde{g}\tilde{\gamma}$. Note that for $\varepsilon
_{o}\rightarrow0$ or $\gamma\rightarrow0$, the integral diverges in the
infrared, indicating that $r_{i,c}\rightarrow-\infty$, i.e. that
$L_{i,c}\rightarrow+\infty$. For $\varepsilon_{o}>0$, however, the term
$\gamma\omega^{1-\varepsilon_{o}}$ introduces a soft cutoff regularizing the
integral. This term dominates below a crossover frequency $\omega^{\ast
}(\varepsilon_{0})=\gamma^{1/\varepsilon_{o}}$, yielding%
\begin{equation}
r_{i,c}(\varepsilon_{o})=-\frac{u}{\pi\varepsilon_{o}}\ln\left(
1/\gamma\right)  .
\end{equation}
The resulting concentration of frozen droplets is $n_{fr}(\varepsilon_{o}%
)\sim\exp\{-\rho r_{i,c}(\varepsilon_{o})/r\}$. These frozen droplets produce
a superparamagnetic contribution to the average susceptibility, proportional
to their concentration $\delta\chi_{fr}\sim n_{fr}(\varepsilon_{o})/T.$ The
ordering temperature can be immediately estimated using arguments similar to
those of Ref.~\cite{smearingprl}. In particular, the RKKY interaction between
droplets falls off as $J(R_{ij})\sim R_{ij}^{-d}$ \cite{rkkytemp}, where the
typical distance between droplets $R_{ij}\sim\left[  n_{fr}(\varepsilon
_{o})\right]  ^{-1/d}$, and we obtain a simple exponential \cite{double-exp}
dependence of $T_{c}$ on the dissipation exponent $\varepsilon_{o}$
\begin{equation}
T_{c}(\varepsilon_{o})\sim n_{fr}(\varepsilon_{o})\sim\exp\left\{  -\frac{\rho
u}{\pi\left\vert r\right\vert \varepsilon_{o}}\ln\left(  1/\gamma\right)
\right\}  . \label{tc}%
\end{equation}

Note that, within our model, the fraction of frozen droplets which
magnetically order vanishes at a \textit{finite} value of the coupling
constant, producing a sharp phase transition from a quantum paramagnet to a
cluster glass phase. In contrast to the infinite randomness fixed point (IRFP)
scenario \cite{dsf9295,senthilsachdev,Motrunich00}, this transition does not
have any percolation-like features. In the itinerant case we consider, even
very distant droplets will order as soon as they freeze, because they interact
through \emph{long-ranged} RKKY interactions. In this case, percolation
processes \cite{senthilsachdev,Motrunich00} associated with the IRFP scenario
are not possible, suggesting that our phase transition is qualitatively
different from those discussed in previous work.

We conclude that the standard QCP scenario is thus qualitatively modified by
the dissipative effects, since an intermediate cluster glass phase generically
emerges between the quantum disordered and any magnetically ordered phase in
itinerant systems. It is easy to see that our conclusions also apply to models
with arbitrary symmetry of the order parameter, since they all find themselves
effectively \textit{above} their lower critical dimension \cite{smearingprl}
due to the presence of non-Ohmic dissipation.

We should emphasize that our finite temperature ordering emerges over the
entire region ($\alpha=d/z^{\prime}<1$) where the Griffiths anomalies arise in
the absence of RKKY interactions (see Fig.~\ref{cap:fig1}). Finite temperature
ordering persists even further (until $\alpha=d/z^{\prime}=2$), but not all
the way to the clean model critical point (corresponding to $\alpha
=d/z^{\prime}\rightarrow+\infty$ according to the large-$N$ estimates of
Ref.~\cite{VojtaSchmalian}). Of course, since the droplets are diluted, the
\emph{magnitude} of the non-Ohmic dissipation produced by the RKKY
interactions is expected to be considerably smaller then the standard Ohmic
dissipation due to Landau damping. As a result, the {}effects of non-Ohmic
dissipation that we predict should be seen only at fairly low temperatures
$T<T^{\ast}\sim\omega^{\ast}=\exp\{-\ln\left(  1/\gamma\right)  /\varepsilon
_{o}\}$, while the Griffiths phase behavior predicted by
Ref.~\cite{VojtaSchmalian} should be observable in a broad interval above this
temperature. Note that for $\rho u>\pi|r|$ (see Eq. \ref{tc}), $T_{c}%
<<T^{\ast}$ for $\varepsilon_{o}<<1$, so there exists an appreciable
temperature window where the the superparamagnetic contribution $\delta
\chi_{fr}\sim1/T$ of frozen droplets should be observable. Otherwise
$T_{c}\approx T^{\ast}$, and the Vojta-Schmalian Griffiths phase should emerge
immediately above the ordering temperature. This scenario, valid for models
with continuous symmetry of the order parameter, should be contrasted to that
expected for the Ising universality class. In that special case, much stronger
Ohmic dissipation is sufficient to suppress the low temperature Griffiths
anomalies. For the Ising case, the existence of an intermediate temperature
range where they would still be observable \cite{CastroNetoJones} remains the
subject of some controversy \cite{moredebate}.

In summary, we have established that the long-ranged part of the RKKY
interaction between the droplets represents a \emph{singular perturbation}
within Griffiths phases. Even when treated to leading order, the RKKY
interactions qualitatively modify the droplet dynamics, leading to droplet
freezing and suppression of conventional quantum critical behavior. Our
arguments strongly suggest that the intervening cluster glass phase arises as
a generic phenomenon \cite{largeD} for itinerant systems with disorder. This
mechanism is, however, specific to those quantum critical points associated
with spontaneous symmetry breaking of a static order parameter. It would be
very interesting to investigate if similar scenarios apply for other quantum
critical points, such as the disorder-driven metal-insulator transition
\cite{mottanderson} and the associated electronic Griffiths phases
\cite{mottanderson,miranda}. This fascinating direction remains a challenge
for future work.

We acknowledge fruitful discussions with E. Abrahams, A. H. Castro Neto, A. J.
Millis, D. K. Morr, J. Schmalian, S. Sachdev, T. Vojta, and K. Yang. This work
was supported by the NSF through grant NSF-0234215 (V.D.), by FAPESP through
grant 01/00719-8 (E.M.), by CNPq through grant 302535/02-0 (E.M.), and the
National High Magnetic Field Laboratory (V.D. and E.M.). We also thank the
Aspen Center for Physics, where part of this work was carried out.


\begin{thebibliography}{99}                                                                                               %


\bibitem {McCoy69}B. M. McCoy, Phys. Rev. Lett. \textbf{23}, 383 (1969).

\bibitem {dsf9295}D. S. Fisher, Phys. Rev. Lett. \textbf{69}, 534 (1992);
Phys. Rev. B \textbf{51}, 6411 (1995).

\bibitem {ThillHuse95}M. Thill and D. Huse, Physica A \textbf{214}, 321 (1995).

\bibitem {YoungRieger96}A. P. Young and H. Rieger, Phys. Rev. B \textbf{53},
8486 (1996).

\bibitem {RiegerYoung96}H. Rieger and A. P. Young, Phys. Rev. B \textbf{54},
3328 (1996).

\bibitem {gbh96}M. Guo \emph{et al}., Phys. Rev. B \textbf{54}, 3336 (1996).

\bibitem {senthilsachdev}T. Senthil and S. Sachdev, Phys. Rev. Lett.
\textbf{77}, 5292 (1996).

\bibitem {Pich98}C. Pich \emph{et al}., Phys. Rev. Lett. \textbf{81}, 5916 (1998).

\bibitem {Motrunich00}O. Motrunich \emph{et al}., Phys. Rev. B \textbf{61},
1160 (2000).

\bibitem {miranda}D. Tanaskovi\'{c} \emph{et al}., cond-mat/0405005; E.
Miranda and V. Dobrosavljevi\'{c}, Phys. Rev. Lett. \textbf{86}, 264 (2001);
E. Miranda \emph{et al}., Phys. Rev. Lett. \textbf{78}, 290 (1997).

\bibitem {CastroNetoJones}A. H. Castro Neto \emph{et al}., Phys. Rev. Lett.
\textbf{81}, 3531 (1998); A. H. Castro Neto and B. A. Jones, Phys. Rev. B
\textbf{62}, 14975 (2000).

\bibitem {MillisMorrSchmalian}A. J. Millis \emph{et al}., Phys. Rev. Lett.
\textbf{87}, 167202 (2001); Phys. Rev. B \textbf{66}, 174433 (2002).

\bibitem {smearingprl}T. Vojta, Phys. Rev. Lett. \textbf{90}, 107202 (2003).

\bibitem {VojtaSchmalian}T. Vojta and J. Schmalian, cond-mat/0405609.

\bibitem {threebody}The formal procedure of integrating out the {}``bulk''
separating droplets will, rigorously speaking, also generate effective
interactions between three and more droplets. However, in the regime where the
density of droplets is exponentially small, these contributions represent
sub-leading terms, which can be safely neglected.

\bibitem {narozhny}B. Narozhny \emph{et al}., Phys. Rev. B \textbf{62}, 14898 (2000).

\bibitem {jagannathan}A. Jagannathan \emph{et al}., Phys. Rev. B \textbf{37},
436 (1988).

\bibitem {sachdev}S. Sachdev, \emph{Quantum Phase Transitions}, (Cambridge
University Press, 2001).

\bibitem {kosterlitz76}J. M. Kosterlitz, Phys. Rev. Lett. \textbf{37}, 1577 (1976).

\bibitem {rkkytemp}We ignore complications due to the \emph{finite
temperature} exponential cutoff of the RKKY power law.

\bibitem {double-exp}Ref.~\cite{smearingprl} suggested a double exponential
form, in contrast to what we find. This follows from assuming that the
droplet-droplet interactions decrease \emph{exponentially} with distance, as
one expects for insulating magnets in a paramagnetic phase. For itinerant
systems, in contrast, the magnetic RKKY interactions are generically long
ranged, giving rise to a simple exponential dependence.

\bibitem {moredebate}A. H. Castro Neto and B. A. Jones, cond-mat/0411197 and
cond-mat/0412020; A. J. Millis, D. K. Morr, and J. Schmalian, cond-mat/0411738.

\bibitem {largeD}The droplet arguments that we and others have used to discuss
the Griffiths phase behavior in principle apply in all spatial dimensions. Our
self-consistent approach to RKKY interactions, however, is rigorously
justified only for sufficiently large dimensions. The role of spatial
fluctuations in low dimensions and the possible modifications of our scenario
in this limit are subtle questions that require further work.

\bibitem {mottanderson}V. Dobrosavljevi\'{c} and G. Kotliar, Phys. Rev. Lett.
\textbf{78}, 3943 (1997); V. Dobrosavljevi\'{c} \emph{et al.,} Phys. Rev.
Lett. \textbf{90}, 016402 (2003).
\end{thebibliography}
\end{document}